\documentclass[conference]{IEEEtran}
\usepackage{cite}
\usepackage{amsmath,amssymb,amsfonts}
\usepackage{algorithm}
\usepackage{algpseudocode}
\usepackage{graphicx}
\usepackage{textcomp}
\usepackage{xcolor}
\usepackage{subcaption}
\usepackage{makecell}
\usepackage{stfloats}
\usepackage[hidelinks]{hyperref}
\def\BibTeX{{\rm B\kern-.05em{\sc i\kern-.025em b}\kern-.08em
    T\kern-.1667em\lower.7ex\hbox{E}\kern-.125emX}}

\begin{document}

\tolerance=1000
\emergencystretch=3em
\hyphenpenalty=10000

\title{System-Wide Termination in Distributed Betweenness Centrality Computation}

\author{
\IEEEauthorblockN{Siamak Abdi}
\IEEEauthorblockA{
Faculty of Engineering\\
Free University of Bozen-Bolzano\\
Bolzano, Italy\\
siamak.abdi@unibz.it}
\and
\IEEEauthorblockN{Lucia Cavallaro}
\IEEEauthorblockA{
Institute for Computing and Information Sciences\\
Radboud University\\
Nijmegen, The Netherlands\\
lucia.cavallaro@ru.nl}
\and
\IEEEauthorblockN{Giuseppe Di Fatta}
\IEEEauthorblockA{
Faculty of Engineering\\
Free University of Bozen-Bolzano\\
Bolzano, Italy\\
giuseppe.difatta@unibz.it}
}

\maketitle
\begin{figure*}[!b]
\centering
\begin{minipage}{0.96\textwidth}
\centering
\fontsize{6}{7}\selectfont
979-8-3315-5021-9/26/\$31.00~\textcopyright~2026 IEEE\\[-0.1em]
\textcopyright~2026 IEEE. Personal use of this material is permitted. Permission from IEEE must be obtained for all other uses, in any current or future media, including reprinting/republishing this material for advertising or promotional purposes, creating new collective works, for resale or redistribution to servers or lists, or reuse of any copyrighted component of this work in other works.
\end{minipage}
\end{figure*}
\bstctlcite{IEEEtran:BSTcontrol}
\begin{abstract}
Computing betweenness centrality on large networks is inherently expensive, as it requires aggregating shortest-path dependencies across all pairs of vertices and becomes increasingly difficult to scale as network size grows. Scalable distributed algorithms can facilitate such computations, particularly when centralised processing is not feasible, and message exchanges must be carefully controlled, for example, in bandwidth-limited or very large-scale networks. However, existing distributed betweenness centrality implementations do not integrate a lightweight, system-wide termination detector. As a consequence, this can lead to extra messaging after local convergence or, if misconfigured, premature stops. In this work, a lightweight, system-wide global termination detection algorithm for this task is presented. The proposed method enables vertices to decide locally when the overall system has converged. The method is evaluated against a local stopping strategy in which vertices terminate individually after their own estimates stabilise. To compare these two approaches, namely global termination detection and local stopping, a custom Python simulator is implemented, and both approaches are tested on synthetic (Erd\H{o}s--R\'enyi, and \textit{Geometric}) and real (\textit{Email} and \textit{Road}) network topologies. Our results show that system-wide termination detection lets vertices stop safely after detecting global convergence, as indicated by zero final error in the evaluated networks, rather than stopping independently based only on local convergence. The local stopping approach, on the other hand, results in premature termination and some errors on heterogeneous networks. This work emphasises the need for coordinated halting in distributed centrality computation.
\end{abstract}

\begin{IEEEkeywords}
Betweenness centrality, global termination detection, distributed graph algorithms, scalable network analysis, distributed computing
\end{IEEEkeywords}

\section{Introduction}
\label{sec:introduction}
Centrality metrics such as degree, closeness, betweenness, and eigenvector centrality are widely used in Network Science to identify important, influential, or structurally critical nodes in a network. Their computation, however, becomes challenging in large networks with many nodes (or vertices) and links (or edges), where the time complexity of centralised algorithms can be prohibitively high~\cite{brandes2001betweenness, saxena2020survey}. Two main approaches have been proposed in the literature, namely the centralised and distributed ones. In the first approach, a central vertex performs the calculations and distributes the results to other vertices; however, this approach suffers from a single point of failure and often incurs high communication overhead. 

For example, let us consider a simple 5-vertex path $v_1{-}v_2{-}v_3{-}v_4{-}v_5$, where betweenness centrality measures how often a vertex lies on shortest paths between other pairs of vertices. In a centralised setting, a vertex must collect shortest-path information for all vertex pairs to compute these values. Even in such a tiny graph, the processor needs $\Theta(n^2)$ path information to decide that $v_3$ has the highest betweenness centrality. In contrast, a distributed approach based on Brandes' shortest-path dependency accumulation~\cite{brandes2001betweenness} allows each vertex to exchange only local distance and path-count updates with its neighbours. In this way, vertices exchange per-target summaries sufficient for dependency accumulation, rather than broadcasting full path lists~\cite{crescenzi2020simple}.

Distributed algorithms allow vertices to compute centrality values collaboratively through local message exchanges, without relying on a single central processor. Examples include decentralised two-phase schemes and Bellman--Ford-compatible distributed betweenness algorithms~\cite{lehmann2003decentralized,hua2016nearly,crescenzi2020simple}. However, many such implementations do not include a global termination detector and often rely on fixed rounds or local inactivity to stop. This is a limitation because, in distributed betweenness computation, shortest-path and dependency information may still be propagating through other parts of the network even when a vertex appears locally stable. As a result, premature termination can introduce residual error, whereas delayed termination can lead to unnecessary communication overhead. This challenge is especially relevant in communication-constrained settings, such as the \textit{CONGEST} model, where each message has a bounded size~\cite{hua2016nearly}, and more generally in large-scale networks where message traffic must be carefully controlled.

In this work, we add a termination layer that detects global quiescence or convergence across all vertices to stop the running of the algorithm at the right time. The proposed algorithm is a fully distributed termination detection algorithm that runs in parallel with the main distributed computation of the centrality algorithm. Each vertex stops executing once it detects that all vertices have converged. This detection process is performed using an estimation process. In contrast, under local convergence, each vertex stops as soon as its own betweenness estimate stabilises.

To summarise, the main contributions of this work are as follows: we propose a lightweight global termination detection layer for distributed betweenness centrality computation that can be executed concurrently with a distance-vector–compatible betweenness algorithm. We demonstrate the need for a globally safe termination detector in the presence of local convergence (\textit{i.e.}, each vertex stops when its own estimate stabilises) by showing that local convergence is unsafe on heterogeneous graphs, resulting in non-zero global error due to missing dependency propagation. We then show that our termination detection identifies an explicit globally safe stopping condition, and we evaluate our approach using a distributed, event-driven simulator to quantify how the communication assumption used for termination detection (overlay \textit{vs} physical-neighbour communication) affects the time to global convergence. 

\section{Related Work}
\label{sec:literature}
Previous works on distributed centrality calculation include synchronous message passing methods, algorithms that consider communication restrictions, as well as distance-vector-compatible approaches. This section reviews the works most relevant to distributed betweenness centrality computation and distributed termination detection.

Lehmann and Kaufmann~\cite{lehmann2003decentralized} proposed an early decentralised framework for centrality calculation based on synchronised message exchanges and two phases for shortest-path counting and back-propagation. Hua \textit{et al.}~\cite{hua2016nearly} studied exact distributed betweenness centrality under the \textit{CONGEST} model and showed that strong communication guarantees are possible even under strict message-size constraints.

From a deployment perspective, compatibility with existing network protocols is valuable. Crescenzi \emph{et al.}~\cite{crescenzi2020simple} describe a distributed algorithm to compute betweenness centrality in distributed networks with little modification of existing algorithms, extending the Bellman--Ford distance-vector algorithm. This algorithm is iterative and works by exchanging quadruples $(t,d,s,b)$. Each vertex iteratively calculates the shortest paths to all other vertices. Quadruples are used for communication between vertices, with $t$ being the target vertex, $d$ the distance estimate, $s$ the number of shortest paths, and $b$ the current betweenness centrality contribution. Distance $d$ is set to infinity and $s$ to $0$ for every vertex except the source. Vertices recursively update $d$ and $s$ with the Bellman--Ford algorithm. Neighbours receive a concise record $(t,d,s,b)$ from every vertex and for each target destination $t$: distance to $t$, number of shortest paths $s$, and the current contribution $b$ employed by the bottom-up accumulation~\cite{crescenzi2020simple}.

Because the work of Crescenzi \emph{et al.}~\cite{crescenzi2020simple} is the closest to our setting, we describe their approach in slightly more detail. In their approach, vertices calculate the betweenness contributions in reverse order, starting from the leaf vertices of the shortest path tree. First, a vertex aggregates values received from its successors and adds its contribution. The contribution sent to its predecessor is proportional to the fraction of shortest paths which pass through that vertex and successor, thus ensuring that each vertex correctly accumulates its betweenness centrality score. 

In the distance-vector variant of the algorithm, convergence occurs in $O(Diam)$ Bellman--Ford phases; in unweighted \textit{CONGEST(1)} settings (\textit{i.e.}, the \textit{CONGEST} model with $O(1)$-size messages per edge per round), exact distributed betweenness centrality can be computed in $O(n)$ rounds, with additional care for aggregation and large-value issues~\cite{pontecorvi2018distributed}. The algorithm applies to both directed and undirected graphs, with the undirected case handled by replacing each edge with its two symmetric directed counterparts. This algorithm is also of practical importance to a variety of applications in large-scale networks due to its simplicity and compatibility with real-world distance-vector algorithms, especially for routing and resource optimisation~\cite{crescenzi2020simple}. Our work builds on this approach, but differs from it by adding a separate global termination detection layer to decide when the distributed computation can stop safely.

The classical approaches to termination detection in a distributed system have been widely studied, including Dijkstra and Scholten’s approach~\cite{dijkstra1978termination} using a diffusing computation framework that relies on termination identification through parent-child acknowledgement, and Mattern’s distributed termination detection algorithms~\cite{mattern1987algorithms} that are based on global state and message-accounting techniques. Instead of using a specific structured approach to termination detection, a lightweight aggregation and epidemic-style approach is used that better matches the decentralised communication model of our distributed betweenness computation.

Overall, these works show that distributed centrality can be computed efficiently, in some cases even under strong communication constraints. Among them, the algorithm of Crescenzi \textit{et al.}~\cite{crescenzi2020simple} is the closest to our setting, since it extends a distance-vector-compatible approach and computes exact betweenness centrality with little modification to existing algorithms. However, prior distributed betweenness methods do not address globally safe termination after local convergence, while classical termination-detection algorithms are not integrated with the betweenness computation considered here. Our work addresses this gap by adding a termination layer on top of the computation without changing the core betweenness algorithm. Other exact distributed BC approaches, such as Hua \textit{et al.}~\cite{hua2016nearly} and Pontecorvi and Ramachandran~\cite{pontecorvi2018distributed}, provide stronger theoretical guarantees under different distributed models. However, they are not directly comparable to our distance-vector setting because these works propose complete BC algorithms and focus on reducing the number of communication rounds required for the computation, whereas our work keeps the distance-vector BC computation unchanged and focuses on detecting when it can stop safely.

\section{Background}
\label{sec:background}
The relevant background is presented in this section. In particular, the definition of betweenness centrality and the distributed betweenness centrality algorithm introduced by Crescenzi \textit{et al.}~\cite{crescenzi2020simple} are described, as they form the basis for our work.

In a graph (or network) $G=(V, E)$, $V$ denotes the set of vertices (or nodes) and $E$ denotes the set of edges (or links); vertices do not have self-loops, and there is at most one directed/undirected edge between any two vertices $u$ and $v$. The number of vertices is $|V|$, but is also denoted by $N$ when referring to the system size in termination conditions.

Once the graph structure is defined, different centrality metrics can be used to quantify the structural importance of vertices. Among these metrics, betweenness centrality is one of the most important and computationally demanding, and has therefore received significant attention in distributed settings. The \emph{betweenness centrality} of a vertex $v$, denoted as $C_B(v)$, is based on the fraction of shortest paths between pairs of vertices that pass through $v$, and is defined in Eq~\ref{eq:cb} as:
\begin{equation}\label{eq:cb}
C_B(v) = \sum_{s \in V} \sum_{\substack{t \in V \\ t \neq s}} P_{st}(v)
\end{equation}
where $s$ and $t$ are two distinct vertices in $V$, $P_{st}(v) = \frac{\sigma_{st}(v)}{\sigma_{st}}$, $\sigma_{st}$ is the total number of shortest paths between $s$ and $t$, and $\sigma_{st}(v)$ is the number of such shortest paths that pass through $v$. \\

Crescenzi \textit{et al.}, utilising the Distributed Bellman--Ford algorithm~\cite{Bellman1958,BertsekasGallager1992}, introduced a fast distributed computation of betweenness centrality, as it allows shortest-path information to be propagated efficiently across the network. The algorithm uses the Bellman--Ford algorithm to find the number of shortest paths among the vertices. Every vertex $v$ is equipped with a state which keeps record of changes of $D[t], NH[t], PH[t], B[u,t], S[u,t]$, where $D[t]$ stores the distance from vertex $v$ to a target vertex $t$, $NH[t]$ represents the next hop from vertex $v$ to the target vertex $t$, and $PH[t]$, similarly, represents the previous hop from the target vertex $t$ to vertex $v$. $B[u,t]$ stores the converged betweenness values of vertex $u$, where $u$ is a physical neighbour of $v$, to the target vertex $t$ and $S[u,t]$ stores the number of shortest paths from vertex $u$ to the target vertex $t$. In the initialisation phase of the algorithm, every vertex sets its $S[v,v]$ to $1$ and $D[v]$ to $0$. We refer the reader to Crescenzi \textit{et al.} work for the full pseudocode. This distributed betweenness centrality algorithm is the baseline algorithm considered in this work.

The algorithm is run synchronously by the vertices based on a defined phase structure. During every phase, every vertex sends a message to all its physical neighbours and receives a message from its neighbours, so then it updates its state. Every vertex updates its own betweenness estimate based on the relationship described in Eq.~\ref{eq:sum-B} after receiving a message from a neighbour. 

\begin{equation}
    \label{eq:sum-B}
    C \gets \displaystyle \sum_{x \neq v} B[v,x]
\end{equation}

The authors define the stop condition for the simulator when all vertices reach convergence on their betweenness estimates. However, this provides a stop condition to halt the simulation; vertices do not know when to stop running the algorithm. While they provide an upper bound on the number of synchronous phases until termination, using such a bound as a termination condition assumes that all vertices know a common global upper bound on the network diameter. This is where our approach differs, as we do not make this assumption, because the aim is to detect termination through global convergence, instead.

Crescenzi \textit{et al.} also introduce an enhanced version of their algorithm (see Algorithm 3 of their paper), which utilises an auxiliary array $A[u,t]$ that stores the intermediate calculation of the betweenness estimate at every time, so vertices do not need to calculate the related equation again. They only update the result of the calculation~\ref{eq:sum-B-over-C} in $A[u,t]$, decreasing the running complexity of the algorithm. 

\begin{equation}
    \label{eq:sum-B-over-C}
    \displaystyle \sum_{x \in PH[t]} \frac{B[x,t] + 1}{S[x,t]}
\end{equation}

As said, this algorithm provides the computational basis for our work. However, it does not by itself address globally safe termination, which motivates the approach presented next.

\section{Methodology}
\label{sec:methodology}
This section details the proposed global termination detection algorithm and explains how it works together with the baseline distributed betweenness centrality computation considered in this paper. Herein, the term \emph{termination layer} refers to the additional distributed mechanism that runs in parallel with the baseline betweenness computation and detects when the whole system has reached a globally converged state.

Although the distributed betweenness centrality computation follows Crescenzi \textit{et al.}~\cite{crescenzi2020simple}, which is herein referred to as the baseline, the proposed termination algorithm is inspired by the agreement and convergence mechanisms introduced in our previous works~\cite{ayiad2016agreement,abdi2025blockchain,abdi2025fully}. It is a fully distributed termination detection algorithm that provides the global information needed for safe stopping. Vertices may reach local convergence at different phases. If some vertices stop as soon as their own estimates stabilise, they may stop before dependency information has fully propagated through the network, which can lead to incorrect betweenness values at other vertices. For this reason, the proposed global termination detection algorithm allows vertices to stop only after the convergence information has been aggregated at the system level.

In our proposed algorithm, every vertex is equipped with a pair of values of $v$ and $w$, which are initialised with $v=0$ and $w=0$ for all vertices, except $v=0$ and $w=1$ for a seed vertex. This algorithm also works in predefined phases. During each phase, vertices first send a push message to a random peer and reply to a push message received from a neighbour then they receive a pull message from the sent neighbour. Whenever a vertex's BC estimate remains within the local stability tolerance for \texttt{MIN} consecutive phases, it enters \texttt{LOCAL CONVERGENCE} and increments $v$ by $1$. It continues running the baseline BC algorithm until \texttt{GLOBAL CONVERGENCE} is detected. We use \texttt{LOCAL CONVERGENCE} to denote vertex-level stabilisation of the betweenness estimate, and \texttt{GLOBAL CONVERGENCE} to denote the system-wide termination condition, which is detected when the aggregated estimate $v/w$ remains within $\epsilon$ of $N$ for at least \texttt{MIN} consecutive phases. Algorithm~\ref{alg:Termination} describes the proposed Global Termination Detection algorithm.

\begin{algorithm}
\caption{Global Termination Detection}
\label{alg:Termination}
\begin{algorithmic}[1]
\Require $N$, $\varepsilon$, \textsc{MIN}, $\textsc{stable}_i$
\Procedure{Init}{$i$}
  \State $v \gets 0$
  \State $w \gets 0$
  \State $activity \gets \textbf{true}$
  \State $state \gets \textsc{RUNNING}$
  \If{seed}
      \State $v \gets 0$
      \State $w \gets 1$
  \EndIf
\EndProcedure
\Procedure{phaseAtvertex}{$i$} \Comment{executed only if $activity=\textbf{true}$}
  \State $j \gets \textsc{GetRandomTarget}()$
  \State $v \gets v/2$
  \State $w \gets w/2$
  \State \textsc{Send}$(j,(v,w),\textit{reply}=\textbf{true})$ \Comment{Push}
  \If{$state == \textsc{RUNNING}$}
    \If{$\textsc{stable}_i$ holds for at least \textsc{MIN} phases}
      \State $state \gets \textsc{LOCAL\_CONVERGENCE}$
      \State $v \gets v + 1$
    \EndIf
  \EndIf
  \If{$state == \textsc{LOCAL\_CONVERGENCE}$}
    \If{$N>0$ \textbf{and} $w>0$ \textbf{and} $\left|\dfrac{N-v/w}{N}\right|\le \varepsilon$ holds for at least \textsc{MIN} phases}
      \State $state \gets \textsc{GLOBAL\_CONVERGENCE}$
      \State $activity \gets \textbf{false}$
    \EndIf
  \EndIf
\EndProcedure
\Procedure{OnReceived}{$m,j,i$}
  \If{$m.\textit{reply}$}
    \State $v \gets v/2$
    \State $w \gets w/2$
    \State \textsc{Send}$(j,(v,w),\textit{reply}=\textbf{false})$ \Comment{Pull}
  \EndIf
  \State $v \gets v + m.v$
  \State $w \gets w + m.w$
\EndProcedure
\end{algorithmic}
\end{algorithm}

The termination layer adds two constant-size logical messages per active vertex per phase: one push and one pull message. Here, an active vertex is a vertex that has not yet detected \textsc{GLOBAL CONVERGENCE}; vertices that have reached \textsc{LOCAL CONVERGENCE} remain active and continue exchanging messages until the global condition is detected. Global convergence detection is asynchronous across vertices, so their stopping phases may differ slightly. This differs from unsafe local stopping because vertices do not stop only when their own betweenness estimate stabilises. Therefore, in phase $k$, the termination layer adds $2A_k$ logical messages, where $A_k \leq N$ is the number of active vertices. Over $T$ phases, the total overhead is $2\sum_{k=1}^{T}A_k \leq 2NT$ logical messages. Each message contains only the pair $(v,w)$, each vertex stores $O(1)$ additional state, and processing each termination message requires $O(1)$ arithmetic operations.

\subsection{System model and communication assumptions}
In our distributed message-passing system, each vertex maintains local state and communicates with other vertices. The distance-vector–compatible betweenness computation follows the physical communication graph: in each phase, each vertex exchanges only the state required by the underlying algorithm with its physical neighbours. In contrast, termination detection can be instantiated under two alternative communication models, namely overlay and physical-neighbour. We consider both models to compare termination detection over physical links versus a logical overlay that can spread convergence information faster.

In the \textit{physical-neighbour model}, termination-detection layer messages are exchanged only along physical edges, and global information propagates via repeated neighbour-to-neighbour interactions. 

In the \textit{overlay model}, each vertex periodically exchanges the state required for termination detection with a peer selected according to an appropriate sampling rule over a logical overlay network, rather than with all neighbours in the underlying graph; these interactions assume the availability of a routing mechanism. Throughout the paper, an overlay interaction corresponds to a single logical message at the application layer, although its implementation may incur multiple hop-level transmissions depending on the routing path. We therefore report stopping phases to global termination; the hop-level communication cost under the overlay model is deployment-dependent.

Unless otherwise stated, each vertex knows its identifier, its current neighbour set, and the network size $N$ (used only to evaluate the global termination condition). We treat phases as logical iterations of the algorithm; in our simulator, messages are delivered in an event-driven fashion, potentially subject to delays, while the algorithmic logic is executed in discrete steps.

According to the assumptions of the baseline algorithm, each vertex knows the set of vertex identifiers $V$ and maintains a per-target state for every $t\in V$. Messages are exchanged only with physical neighbours and carry quadruples ($t,d,s,b$), where $t$ is the target identifier, $d$ the distance estimate, $s$ the number of shortest paths, and $b$ the current betweenness contribution.

Two algorithms are run in parallel using a predefined phase structure. In each phase, every vertex sends two kinds of messages; first, messages related to the betweenness centrality calculation algorithm, where the vertex sends and receives messages to its physical neighbours, and second, messages related to the global termination detection algorithm, where every vertex sends and receives messages to only a random peer. In the proposed approach, whenever a vertex reaches a local convergence, it increments $v$ by $1$ in Algorithm~\ref{alg:Termination}, and when the estimation values of $v/w$, the number of vertices have reached local convergence, match the total number of vertices $N$, it stops running the baseline algorithm. Therefore, at this point, we define global convergence for vertices where all of them have correctly calculated their betweenness centralities.

\subsection{Correctness intuition and limitations}
The proposed termination layer aims to identify a globally safe condition for the distributed betweenness computation, \textit{i.e.}, a point at which no further message exchange affects the final betweenness estimates. This follows from the observation that local convergence alone is not sufficient: when vertices stop independently, dependency information may fail to propagate through parts of the network, freezing the computation in an inconsistent global state.

Each vertex maintains a local indicator of whether it has converged (in the sense that its betweenness estimate has stabilised). Here, ‘stabilised’ means that a vertex’s betweenness estimate does not change by more than a small tolerance for a bounded number of phases. Repeated pairwise exchanges of this indicator—either between physical neighbours or via a peer-sampling overlay—diffuse and aggregate the information. A termination decision is triggered when the aggregated estimate remains close to the known network size $N$ for at least $MIN$ consecutive phases. 

The mechanism targets stable networks under reliable message delivery and a fixed known value of $N$. Transient message delays, including those caused by limited bandwidth or temporary network congestion, only postpone the dissemination of convergence information and therefore delay termination detection. Persistent message loss or node churn are not handled because they may make the aggregated estimate inaccurate. Future extensions of this study might include handling these cases that would require retransmission mechanisms, membership updates, and dynamic updates to $N$. Moreover, under the physical-neighbour communication model, diffusion can be slow on networks with large diameter or strong community structure.

\section{Analysis and Experimental Results}
\label{sec:analysis}
This section evaluates the proposed global termination detection layer on top of the distributed betweenness algorithm in~\cite{crescenzi2020simple}. We developed a Python-based, event-driven simulator of a distributed message-passing system with tunable message and processing delays. This preliminary setup was selected to study the stopping behaviour under controlled delays and communication models, following the baseline algorithm in~\cite{crescenzi2020simple} and our previous epidemic aggregation works~\cite{ayiad2016agreement,abdi2025blockchain,abdi2025fully}. The source code and additional experimental figures are publicly available in the project repository.\footnote{\url{https://github.com/abdi-siamak/distributed_bc_simulation}} Thus, validation on real distributed frameworks, such as Apache Spark/GraphX or MPI, including actual runtime latency and message overhead, is left for future work.

\begin{figure*}
    \centering
    \begin{subfigure}{0.47\textwidth}
        \centering
        \includegraphics[width=\linewidth]{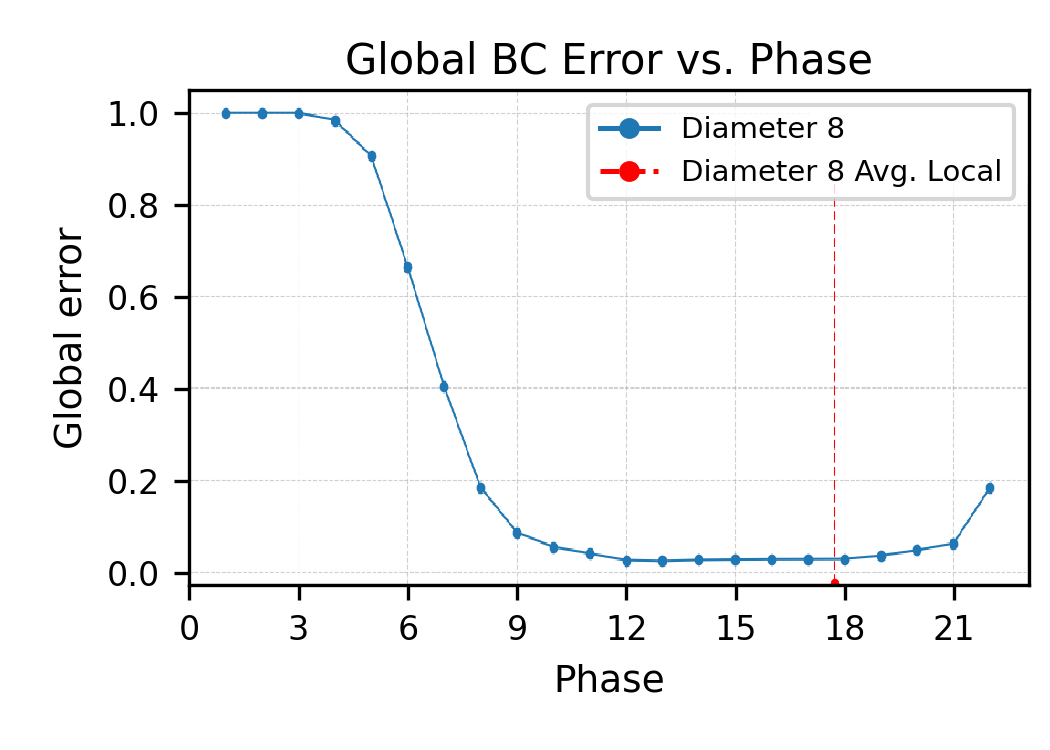}
        \caption{Local convergence termination}
        \label{fig:global-bc-error-er-local}
    \end{subfigure}
    \hfill 
    \begin{subfigure}{0.47\textwidth}
        \centering
        \includegraphics[width=\linewidth]{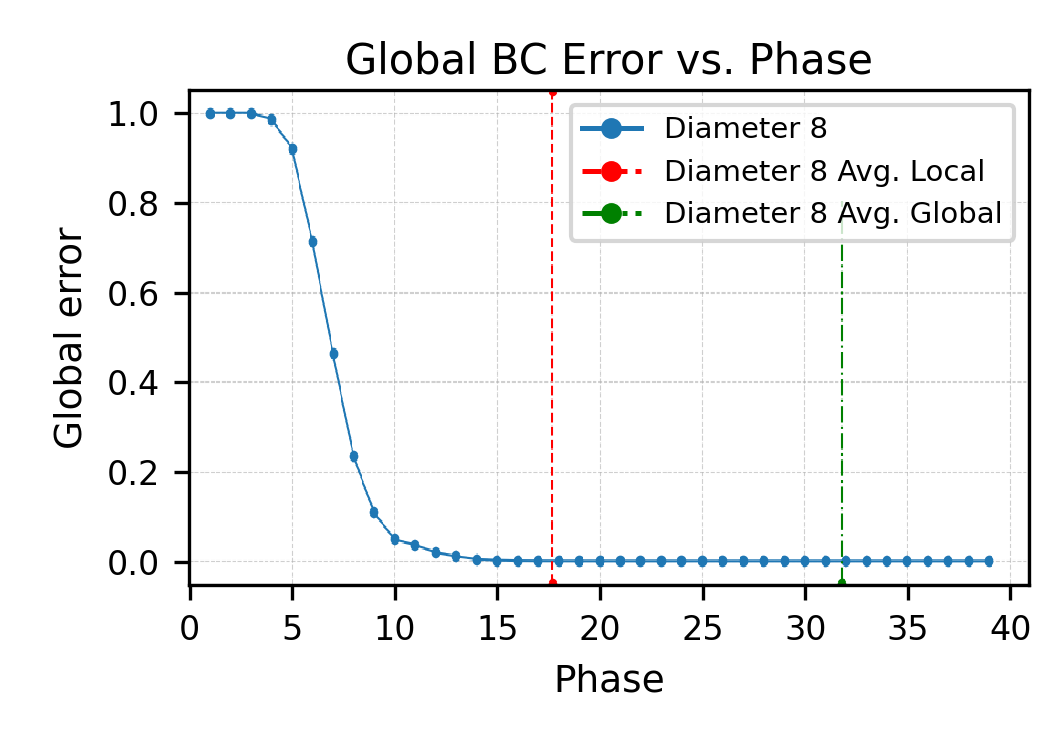}
        \caption{Global convergence termination}
        \label{fig:global-bc-error-er-global}
    \end{subfigure}
    \caption{Comparison of global betweenness centrality error between local and global scenarios for \textit{ER} graph}
    \label{fig:global-bc-error-er}
\end{figure*}

\begin{figure*}
    \centering
    \begin{subfigure}{0.47\textwidth}
        \centering
        \includegraphics[width=\linewidth]{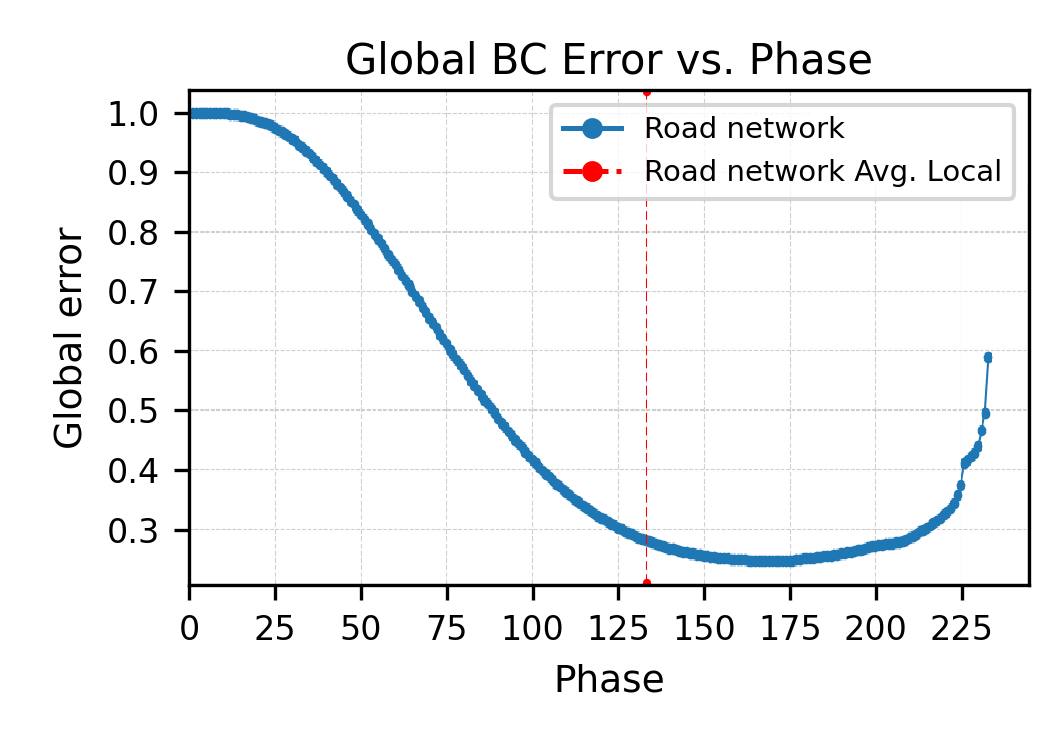}
        \caption{Local convergence termination}
        \label{fig:global-bc-error-road-local}
    \end{subfigure}
    \hfill
    \begin{subfigure}{0.47\textwidth}
        \centering
        \includegraphics[width=\linewidth]{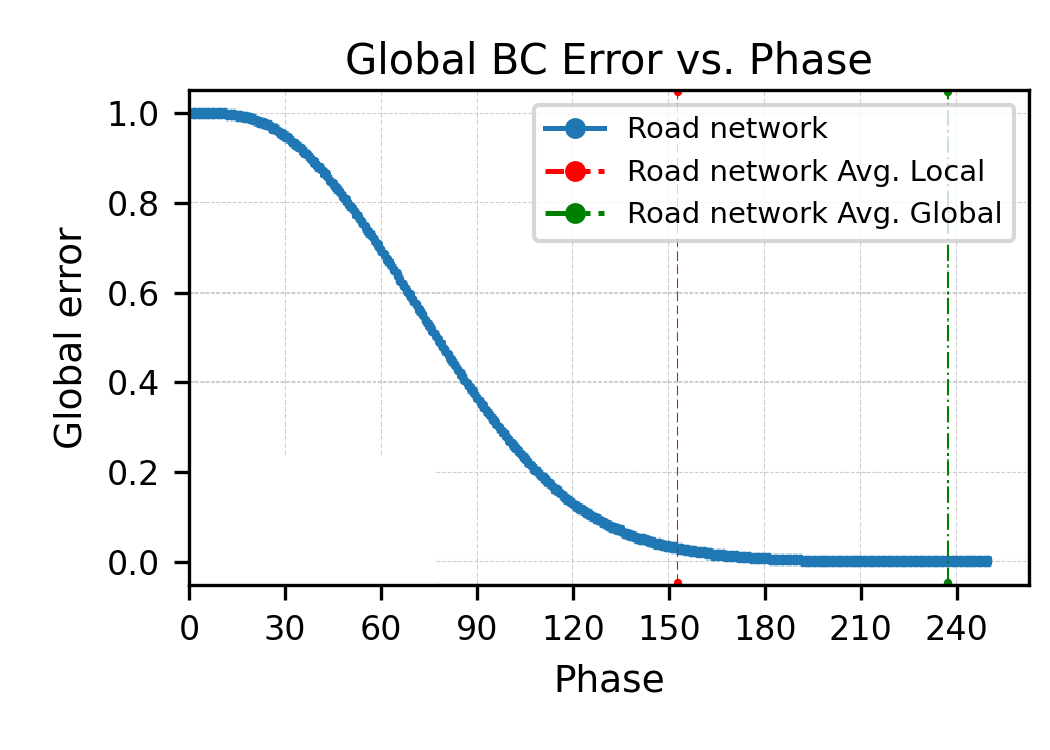}
        \caption{Global convergence termination}
        \label{fig:global-bc-error-road-global}
    \end{subfigure}
    \caption{Comparison of global betweenness centrality error between local and global scenarios for \textit{Road} graph}
    \label{fig:global-bc-error-road}
\end{figure*}

\begin{figure*}
    \centering
    \begin{subfigure}{0.47\textwidth}
        \centering
        \includegraphics[width=\linewidth]{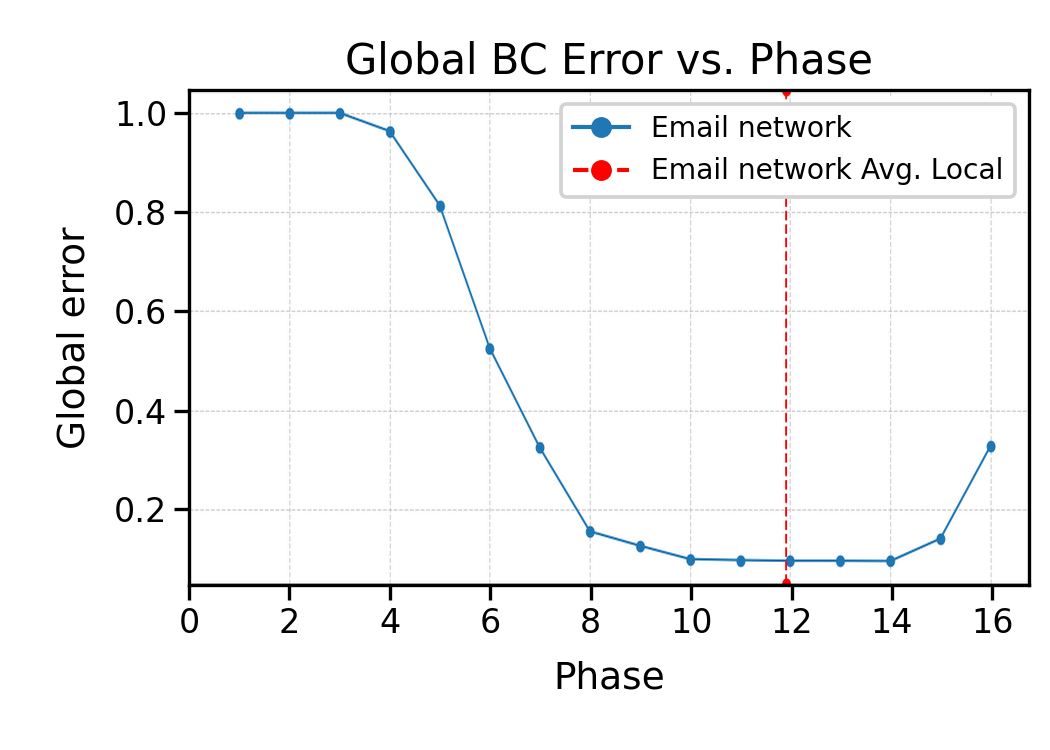}
        \caption{Local convergence termination}
        \label{fig:global-bc-error-email-local}
    \end{subfigure}
    \hfill
    \begin{subfigure}{0.47\textwidth}
        \centering
        \includegraphics[width=\linewidth]{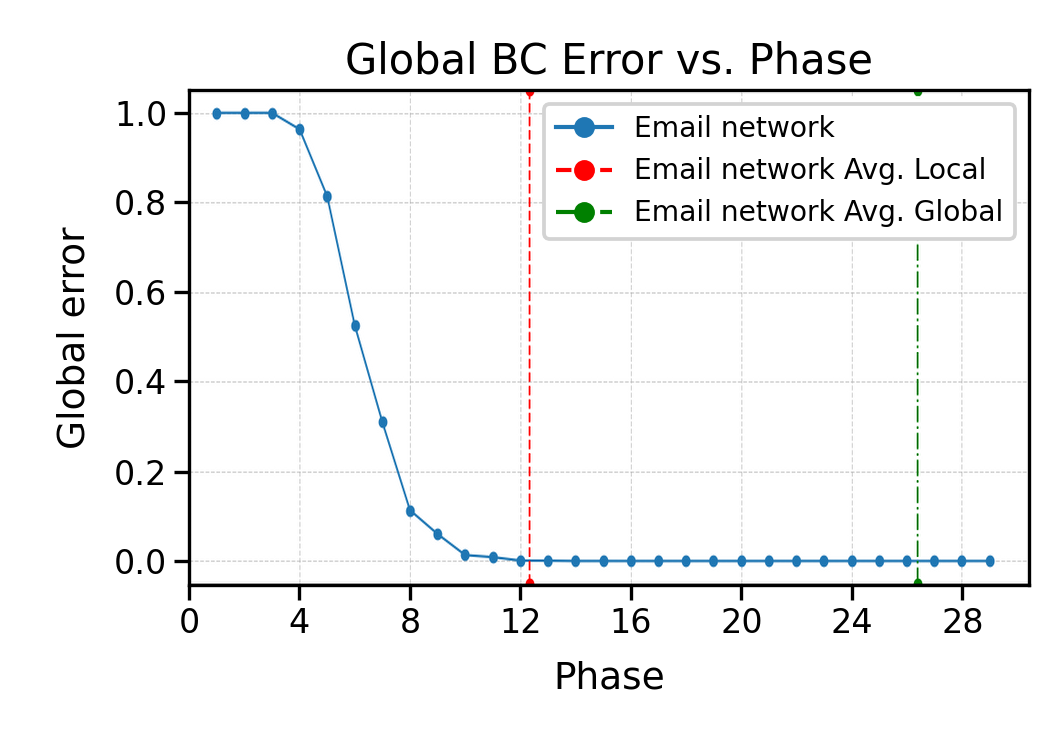}
        \caption{Global convergence termination}
        \label{fig:global-bc-error-email-global}
    \end{subfigure}
    \caption{Comparison of global betweenness centrality error between local and global scenarios for the \textit{Email} graph}
    \label{fig:global-bc-error-email}
\end{figure*}

\begin{figure*}
    \centering
    \begin{subfigure}{0.47\textwidth}
        \centering
        \includegraphics[width=\linewidth]{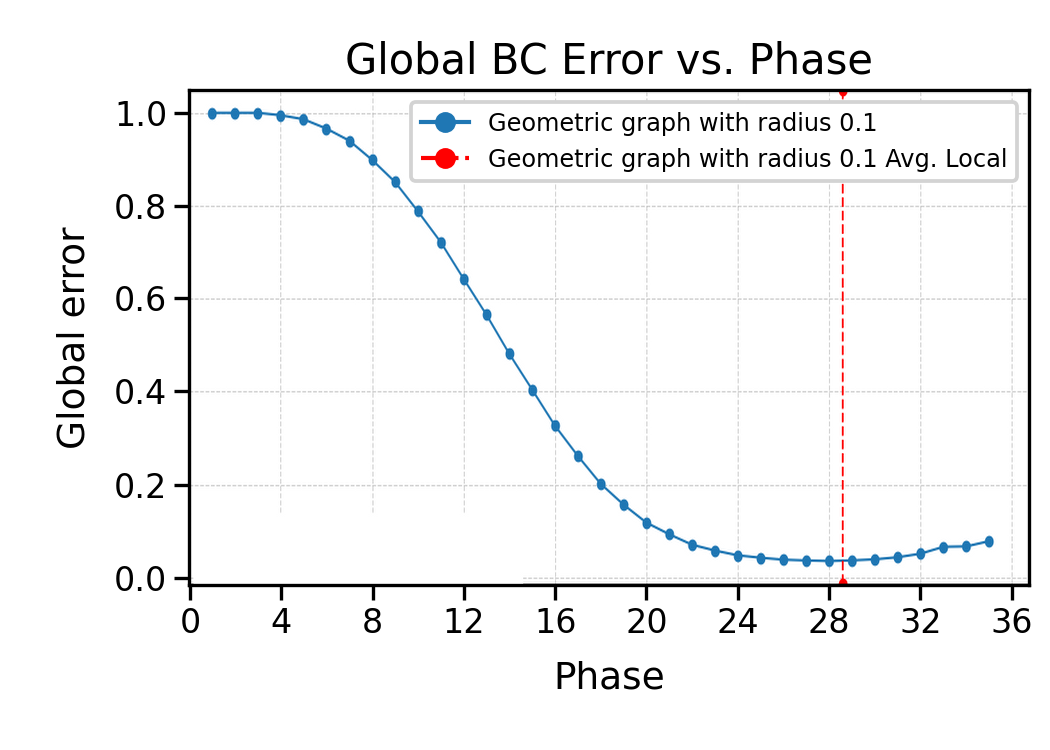}
        \caption{Local convergence termination}
        \label{fig:global-bc-error-geometric-local}
    \end{subfigure}
    \hfill
    \begin{subfigure}{0.47\textwidth}
        \centering
        \includegraphics[width=\linewidth]{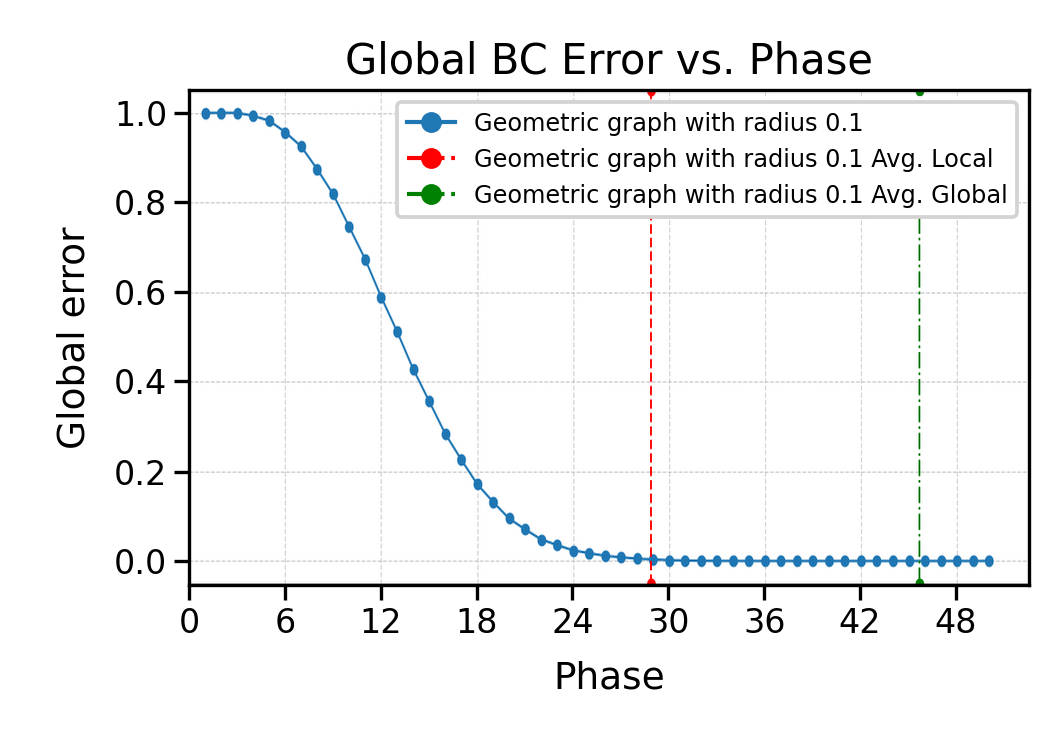}
        \caption{Global convergence termination}
        \label{fig:global-bc-error-geometric-global}
    \end{subfigure}
    \caption{Comparison of global betweenness centrality error between local and global scenarios for the \textit{Geometric} graph}
    \label{fig:global-bc-error-geometric}
\end{figure*}

The processing time was set to $0$\,s, the per-message delay to $100$\, ms, and the phase period to $1$\,s. We set $\epsilon = 0.05$ and \textsc{MIN}$=5$. A vertex was considered locally stable when its BC estimate changed by less than $\epsilon$ for \textsc{MIN} consecutive phases. We considered two stopping scenarios introduced in Section~\ref{sec:methodology}: \emph{local termination}, in which a vertex stops when its own betweenness estimate stabilises, and \emph{global termination}, in which vertices continue until system-wide convergence is signalled by the termination layer. Once a vertex stops, it does not transmit or process messages to/from neighbours. We considered four graphs: two synthetic (Erd\H{o}s--R\'enyi, or \textit{ER} in short, with discrete weights having a diameter $\approx 8$, and \textit{Geometric} with connection radius 0.1), and two real (\textit{Email} with 1,133 vertices and 5,451 edges, and \textit{Road} with 2,000 vertices and 3,500 edges). The synthetic graphs were generated with NetworkX, while the real graphs were taken from publicly available benchmark datasets~\footnote{For the evaluation using real graph datasets, we followed Crescenzi et al.~\cite{crescenzi2020simple} and used the same benchmark graphs, namely the Email network and the Rome 1999 Road network.}. For the termination layer, we considered the two communication models, namely \emph{physical-neighbour} and \emph{overlay}, described in Section~\ref{sec:methodology}. The baseline BC algorithm always uses physical-neighbour communication. The results in Figs.~\ref{fig:global-bc-error-er}--\ref{fig:global-bc-error-geometric} and Table~\ref{tab:stop_efficiency} use the overlay communication model for the termination layer.

Figures~\ref{fig:global-bc-error-er}--\ref{fig:global-bc-error-geometric} illustrate the global betweenness error, computed using the relative $\ell_2$ error in Eq.~\ref{eq:bc-error}, as in~\cite{crescenzi2020simple}. This metric measures the relative $\ell_2$ error between the distributed estimates $\hat{C}_B(v)$ and the ground-truth betweenness centrality values $C_B(v)$ computed using \textit{NetworkX}~\footnote{One of the most widely used Python packages for the creation, manipulation, and study of complex networks: \url{https://networkx.org/en/}.}.

\begin{equation}
\label{eq:bc-error}
\frac{\|\hat{C}_B - C_B\|_2}{\|C_B\|_2}
=
\frac{\sqrt{\sum_{v \in V} \left(\hat{C}_B(v) - C_B(v)\right)^2}}
{\sqrt{\sum_{v \in V} \left(C_B(v)\right)^2}}
\end{equation}
Here, $\hat{C}_B(v)$ denotes the final distributed estimate of the betweenness centrality of vertex $v$, while $C_B(v)$ denotes the corresponding ground-truth betweenness centrality value.


Figures~\ref{fig:global-bc-error-er-local},~\ref{fig:global-bc-error-road-local},~\ref{fig:global-bc-error-email-local}, and~\ref{fig:global-bc-error-geometric-local} illustrate the global betweenness error under the \textit{local termination} scenario. As shown in Figs.~1(a), 2(a), 3(a), and 4(a), the error rises when vertices stop early after reaching local convergence. The explanation is that early-stopped vertices no longer contribute to shortest-path counting or dependency back-propagation, while their neighbours may still require this information to finalise their betweenness estimates. The red dotted line represents the \emph{average} phase of local convergence and depicts that some vertices stop earlier than others.

On the other hand, Figures~\ref{fig:global-bc-error-er-global},~\ref{fig:global-bc-error-road-global},~\ref{fig:global-bc-error-email-global}, and~\ref{fig:global-bc-error-geometric-global} present the results for the \textit{global termination} scenario in which vertices stop only after reaching the global convergence criterion defined by our global termination detection approach. The green dashed line indicates the \emph{average} global convergence phase, beyond which vertices stop. As shown in Figs.~1(b), 2(b), 3(b), and 4(b), the error remains at zero or returns to zero because early-converged vertices remain active between the red and green lines, supplying the remaining shortest-path counts and dependency contributions that some other vertices need to finalise their betweenness estimates.

Collectively, the results in Figures~\ref{fig:global-bc-error-er-global},~\ref{fig:global-bc-error-road-global},~\ref{fig:global-bc-error-email-global}, and~\ref{fig:global-bc-error-geometric-global} demonstrate that the global termination detection algorithm provides an explicit system-wide termination condition and enables vertices to stop safely after detecting global convergence. This prevents premature termination and the resulting betweenness centrality error. Without a global stopping condition, vertices may either run longer than required—experiencing unnecessary message exchanges—or they terminate early, disrupting in-progress dependency accumulation.

Although some vertices may stop earlier under local convergence, Table~\ref{tab:stop_efficiency} shows that stopping at local convergence is unsafe across all evaluated graphs. In all scenarios, local convergence occurs earlier but yields a non-zero error, whereas global convergence occurs later and preserves zero error. Table~\ref{tab:stop_efficiency} summarises the local and global convergence processes and their residual betweenness errors; for exact betweenness, even small non-zero error matters.

\begin{table}
\centering
\caption{Stopping efficiency and accuracy of local \textit{vs}\ global convergence, where $\Delta\text{phases}=T_{\text{global}}-T_{\text{local}}$}
\label{tab:stop_efficiency}
\begin{tabular}{lccccc}
\hline
Graph & \makecell{Local-stop\\phase} & \makecell{Global-stop\\phase} & \makecell{$\Delta$\\phases} & \makecell{Error\\@local} & \makecell{Error\\@global} \\
\hline
Email     & 12  & 27  & 15  & 0.34 & 0.00 \\
ER        & 18  & 33  & 15  & 0.13 & 0.00 \\
Geometric & 28  & 46  & 18  & 0.09 & 0.00 \\
Road      & 130 & 240 & 110 & 0.60 & 0.00 \\
\hline
\end{tabular}
\end{table}

A fixed-round execution that is sufficiently large to account for worst-case convergence may continue after the network has effectively reached a stable state. The proposed detector provides an explicit stopping point once global convergence is detected, while preventing the premature-termination error observed under local convergence. Under the \textit{overlay} communication model, the detector typically converges faster than under the \textit{physical-neighbour} communication model.

In the \textit{physical-neighbour} communication model, more phases were required to reach termination than in the overlay model for every simulated graph, and for the \textit{Geometric} graph, the termination layer did not converge within the given simulation budget.


\section{Conclusion}
\label{sec:conclusion}
In this paper, we introduced a global termination detection layer for distributed betweenness centrality based on a distance–vector (Bellman--Ford) compatible algorithm. Using an event-driven simulator, we showed that relying on local convergence alone leads to premature termination and residual error because early-stopped vertices no longer provide information required by other vertices. In contrast, the proposed algorithm enables vertices to detect global convergence and stop safely, preserving zero final error across all evaluated graphs. The overlay communication model also accelerates termination detection compared with the physical-neighbour communication model, especially for slowly mixing or high-diameter networks. Given the promising results achieved, future works include the validation of the proposed approach on Spark/GraphX or MPI, measure runtime and hop-level communication overhead under congestion and failures, and study its integration with other exact distributed BC approaches.

\section*{Acknowledgment}
This research has been partially supported by the Province of Bolzano and EU through project ERDF-FESR~1047 AI-Lab.
\bibliographystyle{IEEEtran}
\bibliography{bibliography.bib}

@IEEEtranBSTCTL{IEEEtran:BSTcontrol,
  CTLdash_repeated_names = "no"
}

@book{lehmann2003decentralized,
  title={Decentralized algorithms for evaluating centrality in complex networks},
  author={Lehmann, Katharina A and Kaufmann, Michael},
  year={2003},
  publisher={Universit{\"a}tsbibliothek T{\"u}bingen T{\"u}bingen, Germany}
}

@inproceedings{hua2016nearly,
  title={Nearly optimal distributed algorithm for computing betweenness centrality},
  author={Hua, Qiang-Sheng and Fan, Haoqiang and Ai, Ming and Qian, Lixiang and Li, Yangyang and Shi, Xuanhua and Jin, Hai},
  booktitle={2016 IEEE 36th International Conference on Distributed Computing Systems (ICDCS)},
  pages={271--280},
  year={2016},
  organization={IEEE}
}

@inproceedings{crescenzi2020simple,
  title={Simple and fast distributed computation of betweenness centrality},
  author={Crescenzi, Pierluigi and Fraigniaud, Pierre and Paz, Ami},
  booktitle={IEEE INFOCOM 2020-IEEE Conference on Computer Communications},
  pages={337--346},
  year={2020},
  organization={IEEE}
}

@inproceedings{ayiad2016agreement,
  title={Agreement in epidemic information dissemination},
  author={Ayiad, Mosab and Katti, Amogh and Di Fatta, Giuseppe},
  booktitle={Internet and Distributed Computing Systems: 9th International Conference, IDCS 2016, Wuhan, China, September 28-30, 2016, Proceedings 9},
  pages={95--106},
  year={2016},
  organization={Springer}
}

@inproceedings{abdi2025fully,
  title={Fully Decentralised Consensus for Extreme-scale Blockchain},
  author={Abdi, Siamak and Di Fatta, Giuseppe and Badii, Atta and Fortino, Giancarlo},
  booktitle={2025 IEEE Global Blockchain Conference (GBC)},
  pages={1--8},
  year={2025},
  organization={IEEE}
}

@inproceedings{abdi2025blockchain,
  title={Blockchain Epidemic Consensus for Large-Scale Networks},
  author={Abdi, Siamak and Di Fatta, Giuseppe and Badii, Atta and Fortino, Giancarlo},
  booktitle={2025 7th International Conference on Blockchain Computing and Applications (BCCA)},
  pages={562--569},
  year={2025},
  organization={IEEE},
  doi={10.1109/BCCA66705.2025.11229662}
}

@book{BertsekasGallager1992,
  author    = {Dimitri P. Bertsekas and Robert G. Gallager},
  title     = {Data Networks},
  edition   = {2},
  publisher = {Prentice Hall},
  year      = {1992},
  note      = {See \S5.2.4: Distributed Asynchronous Bellman--Ford Algorithm}
}

@article{Bellman1958,
  author  = {Richard Bellman},
  title   = {On a Routing Problem},
  journal = {Quarterly of Applied Mathematics},
  volume  = {16},
  number  = {1},
  pages   = {87--90},
  year    = {1958},
  doi     = {10.1090/qam/102435}
}

@article{brandes2001betweenness,
  title={A faster algorithm for betweenness centrality},
  author={Brandes, Ulrik},
  journal={Journal of Mathematical Sociology},
  volume={25},
  number={2},
  pages={163--177},
  year={2001},
  publisher={Taylor \& Francis}
}

@article{saxena2020survey,
  title={Centrality Measures in Complex Networks: A Survey},
  author={Saxena, A. and Iyengar, S. and others},
  journal={arXiv:2011.07190},
  year={2020}
}

@article{pontecorvi2018distributed,
  title={Distributed algorithms for directed betweenness centrality and all pairs shortest paths},
  author={Pontecorvi, Matteo and Ramachandran, Vijaya},
  journal={arXiv preprint arXiv:1805.08124},
  year={2018}
}

@article{dijkstra1978termination,
  author  = {Dijkstra, Edsger W. and Scholten, Carel S.},
  title   = {Termination Detection for Diffusing Computations},
  journal = {Information Processing Letters},
  volume  = {11},
  number  = {1},
  pages   = {1--4},
  year    = {1980},
  doi     = {10.1016/0020-0190(80)90021-6}
}

@article{mattern1987algorithms,
  title={Algorithms for distributed termination detection},
  author={Mattern, Friedemann},
  journal={Distributed computing},
  volume={2},
  number={3},
  pages={161--175},
  year={1987},
  publisher={Springer}
}
\end{document}